\documentclass[reprint,aps,prl,twocolumn,superscriptaddress]{revtex4-1}
\usepackage{amsmath,hyperref,nccfoots,tensor}
\hypersetup{colorlinks=true,allcolors=blue}
\renewcommand{\[}{\begin{equation}\begin{aligned}}
\renewcommand{\]}{\end{aligned}\end{equation}}

\def\mM{\mathcal{M}}
\def\mR{\mathcal{R}}
\def\mL{\mathcal{L}}

\def\Ft{\tilde{F}}
\def\Rt{\tilde{R}}
\def\mRt{\tilde{\mR}}

\def\pa{\partial}

\def\ep{\epsilon}

\def\gc[#1,#2,#3]{\tensor{\Gamma}{_{#1#2}^{#3}}}
\def\torsion[#1,#2,#3]{\tensor{T}{_{#1#2}^{#3}}}
\def\contorsion[#1,#2,#3]{\tensor{K}{_{#1#2}^{#3}}}
\def\chris[#1,#2,#3]{\left\{\begin{array}{c}#1 \\#2#3 \end{array}\right\}}

\begin{document}

\title{Duality in Gauge Theory, Gravity and String Theory}
\author{Uri Kol}
\affiliation{Center of Mathematical Sciences and Applications, Harvard University, MA 02138, USA}

\author{Shing-Tung Yau}
\affiliation{Center of Mathematical Sciences and Applications, Harvard University, MA 02138, USA}
\affiliation{Department of Mathematics, Tsinghua University, Haidian District, Beijing 100084, China}

\begin{abstract}
Einstein's theory in the vacuum was recently shown to possess an $SO(2)$ duality invariance, which is broken by coupling to matter. Duality invariance can be restored by enlarging the phase space of the theory to allow for violations of the algebraic Bianchi identity. We show that in cases where the matter content can be understood as a component of the torsion tensor duality can be restored and we compute the corresponding duality current. We consider the case of NS-NS gravity as an example and find that the duality current is given by the divergence of the axion. In the linearized approximation of the low energy heterotic string theory these results imply that duality of the generalized Riemann curvature tensor implements Riemannian, axion-dilaton and electro-magnetic dualities simultaneously. 
\end{abstract}

\maketitle

\Footnotetext{}{E-mail addresses:\\ \href{urikol@fas.harvard.edu}{urikol@fas.harvard.edu} , \href{yau@math.harvard.edu}{yau@math.harvard.edu} .	}

\section{Introduction}

Duality is an important concept in physics and mathematics, encompassing a wide array of phenomena.
It refers to the equivalence between two different descriptions of the same system.
Perhaps the most well-known instance of duality in physics is electromagnetic duality in Maxwell's theory
\[
\mL_{\text{Maxwell}}= \frac{1}{4e^2} F_{\mu\nu}F^{\mu\nu}.
\]
Maxwell's equations of motion and the Bianchi identity
\[\label{MaxwellEqns}
\text{Maxwell's equations:} & \qquad &  \pa_{\mu} F^{\mu\nu} &= 0,  \\
\text{Bianchi Identity:} & \qquad  & \pa_{\mu} \Ft^{\mu\nu}  &=0,
\]
are invariant under the following $SO(2)$ duality rotation
\[\label{EMduality}
\begin{pmatrix}
F_{\mu\nu}  \\ \Ft_{\mu\nu} 
\end{pmatrix}
\longrightarrow
\begin{pmatrix}
F_{\mu\nu} ' \\ \Ft_{\mu\nu} '
\end{pmatrix}
=
\begin{pmatrix}
+\cos \theta & +\sin \theta\\
-\sin \theta & +\cos \theta
\end{pmatrix}
\begin{pmatrix}
F_{\mu\nu}  \\ \Ft_{\mu\nu} 
\end{pmatrix}.
\]
Here $\Ft_{\mu\nu} $ is the spacetime Hodge dual of the field strength $F_{\mu\nu}$, which is defined by
\[\label{SpacetimeDualityOperation}
\Ft_{\mu\nu} = \frac{1}{2} \ep _{\mu\nu\rho\sigma}F^{\rho\sigma}.
\]
When the Maxwell action is coupled to matter fields, the equations of motion and Bianchi identity will be sourced by electric and magnetic currents, respectively. Duality will then rotate the sources into each other.

Note that while the dynamical equations of motion are invariant under duality, the Lagrangian transforms non-trivially \cite{Shnir:2005vvi}
\[
\mL_{\text{Maxwell}}\longrightarrow \cos 2\theta \, \mL_{\text{Maxwell}}
+\sin 2\theta  \, \frac{1}{4e^2}  F_{\mu\nu}\Ft^{\mu\nu}.
\]
However, the second term above is a total derivative
\[
F_{\mu\nu}\Ft^{\mu\nu} = 2 \pa_{\mu}D^{\mu},
\]
where we introduced the \emph{duality current}
\[\label{GaugeDualCurrent}
D^{\mu} = A_{\nu}\Ft^{\mu\nu}.
\]
The dual symmetry of the free Maxwell theory implies the current conservation
\[
 \pa_{\mu}D^{\mu} = 0.
\]

Attempts to realize duality in gravitational theories have a long history \cite{Curtright:1980yk,Hull:2001iu}. Duality in linearized gravity was studied in \cite{Henneaux:2004jw,Bunster:2006rt,Argurio:2009xr}, as well as in theories of supergravity \cite{deWit:2013ija}.
More recently, non-linear realization of duality in the full Einstein theory in the vacuum was established in \cite{Kol:2022bsd}.
Couplings to matter break duality invariance in general. In order to restore duality as a classical symmetry of the dynamical equations of motion one has to source the algebraic Bianchi identity. This can be done by introducing torsion.
In this letter we consider cases in which matter fields can be understood as components of the torsion tensor.
We show that in these cases duality invariance can be restored and we compute the corresponding duality current.

As an example, we consider NS-NS gravity, a theory whose field content includes the metric as well as a dilaton and a Kalb-Ramond two-form.
We then consider a compactification of the heterotic string theory on a Calabi-Yau manifold \cite{Yau:1977ms,Yau:1978cfy} down to four dimensions. The field content of the low energy theory contains, in addition to the NS-NS sector, gauge fields. We show that in the linearized approximation, duality rotation of the generalized Riemann curvature tensor implements gravitational, electro-magnetic and axion-dilaton dualities, all simultaneously.

\section{Duality in Einstein's Gravity}\label{EinsteinDuality}

Let us start by reviewing the results of \cite{Kol:2022bsd} on duality in Einstein's gravity.
In the first order formalism, the Einstein-Hilbert action
\[\label{EHaction}
S_{EH}  =  \frac{1}{16 \pi G }\int_{\mM} \, \Rt _{ab}\wedge \theta^{a} \wedge \theta^{b}. 
\]
is expressed in terms of the curvature 2-form $R^{ab}$ and a non-coordinate basis of 1-forms $\theta^a$ in the flat tangent space. The curvature 2-form is the field strength of the spin connection $\omega^{ab}_{\mu}$
\[
R^{ab} = d \omega^{ab} + \omega^a{}_c\wedge \omega^{cb}
\]
and its components are given by the projection of the Riemann tensor onto the tangent space
\[\label{Riemann}
R^{ab} = \frac{1}{2}R^{ab}{}_{cd} \, \theta^c \wedge \theta^d .
\]
The non-coordinate basis of 1-forms in the tangent space is related to the coordinate basis in spacetime through the vielbeins $e^a{}_{\mu}$
\[\label{nonCoordBasis}
\theta^a = e^a{}_{\mu}dx^{\mu}.
\]
Note that we are using a notation in which tangent space indices are denoted by Latin letters $a,b,c$, while spacetime indices are denoted by Greek letters $\mu,\nu,\sigma$.

In writing the action \eqref{EHaction}, we have defined the \emph{tangent space} Hodge duality operation
\[\label{HodgeTangent}
\Rt _{ab}\equiv (\star R )_{ab} \equiv \frac{1}{2}\ep_{abcd}R^{cd},
\]
which is distinct from the \emph{spacetime} Hodge duality operation \eqref{SpacetimeDualityOperation}.

The Einstein equation and the Bianchi identity can be written using differential forms as
\[\label{GravityEqns}
\text{Einstein's equations:} \qquad & \Rt _{ab} \wedge \theta ^b &=0,\\
\text{Bianchi Identity:} \qquad & R _{ab} \wedge \theta ^b&=0.
\]
In components, and upon projection onto spacetime indices, these equations reproduce the familiar form of the Einstein equation $R_{\mu\nu}-\frac{1}{2}g_{\mu\nu}R=0$ and the algebraic Bianchi identity $R^\mu{}_{[\nu\rho\sigma]}=0$, respectively.

It is now evident that the set of equations in \eqref{GravityEqns} is invariant under the following $SO(2)$ duality rotation
\[\label{dualityOperation}
\begin{pmatrix}
R_{ab} \\ \Rt_{ab}
\end{pmatrix}
\longrightarrow
\begin{pmatrix}
R_{ab}' \\ \Rt_{ab}'
\end{pmatrix}
=
\begin{pmatrix}
+\cos \theta & +\sin \theta\\
-\sin \theta & +\cos \theta
\end{pmatrix}
\begin{pmatrix}
R_{ab} \\ \Rt_{ab}
\end{pmatrix}.
\]
The $SO(2)$ duality rotation \eqref{dualityOperation} is a symmetry of the equations of motion and, consequently, it maps different solutions into each other, as demonstrated in \cite{Kol:2022bsd}.

\section{Coupling to Matter}\label{DualityMatter}

In general, coupling to matter breaks the duality invariance of the theory since it sources the Einstein equation while leaving the Bianchi identity intact. In order to restore duality invariance one needs to source the Bianchi identity as well. That can be done by enlarging the phase space of the theory to include torsion. The resulting theory is the Einstein-Cartan theory, in which a manifold is equipped with an affine connection that is not symmetric
\[
\gc[\mu,\nu,\sigma]=\gc[(\mu,\nu),\sigma]+\gc[[\mu,\nu],\sigma].
\]
The symmetric part of the connection is $\gc[(\mu,\nu),\sigma]$, while its antisymmetric part is called the \emph{torsion} and it is a tensor
\[
\torsion[\mu,\nu,\rho] \equiv -2 \gc[[\mu,\nu],\sigma] \equiv - \gc[\mu,\nu,\sigma]+\gc[\nu,\mu,\sigma] .
\]
The Riemannian manifold is recovered when the torsion is set to zero.
The connection can be decomposed into its Riemannian and non-Riemannian parts as follows
\[\label{connection}
\gc[\mu,\nu,\sigma] = \chris[\sigma,\mu,\nu]+
\contorsion[\mu,\nu,\sigma] ,
\]
where
\[
\chris[\sigma,\mu,\nu]
=\frac{1}{2}g^{\sigma\rho}
\left(
\pa_{\mu} g_{\nu\rho} +\pa_{\nu} g_{\mu\rho} -\pa_{\rho} g_{\mu\nu} 
\right)
\]
are the  \emph{Christoffel symbols} and
\[
\contorsion[\mu,\nu,\sigma]  = \frac{1}{2}g^{\sigma\rho}
\left(  T_{\mu\rho\nu}+T_{\nu\rho\mu} -T_{\mu\nu\rho} \right)
\]
is called the \emph{contorsion tensor}.
The contorsion obeys
\[\label{contorsionProperty}
\contorsion[[\mu,\nu],\sigma] = -\frac{1}{2}\torsion[\mu,\nu,\sigma],
\qquad
K_{\mu\nu\sigma} = - K_{\mu\sigma\nu}.
\]
Note that the metric and the torsion are independent degrees of freedom and that the contorsion depends on both.
Note also that the symmetric part of the connection receives a contribution from the contorsion and therefore depends on the torsion as well.

The generalized Riemann tensor in Einstein-Cartan theory, which we will denote by $\mR_{\mu\nu\rho}{}^{\sigma}$, can be decomposed into its Riemannian part $R_{\mu\nu\rho}{}^{\sigma}$ and torsional contributions
\[
\mR_{\mu\nu\rho}{}^{\sigma} = 
R_{\mu\nu\rho}{}^{\sigma} &+\nabla_{\nu}K_{\mu\rho}{}^{\sigma}
-\nabla_{\mu}K_{\nu\rho}{}^{\sigma}
\\&
+K_{\nu\tau}{}^{\sigma}K_{\mu\rho}{}^{\tau}
-K_{\mu\tau}{}^{\sigma}K_{\nu\rho}{}^{\tau},
\]
where $\nabla$ denotes the Levi-Civita connection.
The Bianchi identity is now sourced by the torsion tensor
\[\label{SourcedBianchi}
\mR_{ab}\wedge \theta^b = DR_a,
\]
where $\mR^{ab} = \frac{1}{2}\mR^{ab}{}_{cd} \, \theta^c \wedge \theta^d$ is the generalized curvature 2-form,
\[
R^a = D\theta ^a = - \frac{1}{2}T_{bc}{}^a \theta^b \wedge\theta^c,
\]
is the torsion 2-form and $D$ is the Lorentz covariant derivative with respect to the spin connection.

Let us now consider the Palatini action
\[\label{PalatiniAction}
S_p  =  \frac{1}{16 \pi G }\int_{\mM} \, \mRt _{ab}\wedge \theta^{a} \wedge \theta^{b}.
\]
Under the following $SO(2)$ duality rotation
\[\label{dualityOperation}
\begin{pmatrix}
\mR_{ab} \\ \mRt_{ab}
\end{pmatrix}
\longrightarrow
\begin{pmatrix}
\mR_{ab}' \\ \mRt_{ab}'
\end{pmatrix}
=
\begin{pmatrix}
+\cos \theta & +\sin \theta\\
-\sin \theta & +\cos \theta
\end{pmatrix}
\begin{pmatrix}
\mR_{ab} \\ \mRt_{ab}
\end{pmatrix}
\]
the Palatini action transforms as
\[
S_p \longrightarrow \cos \theta \, S_p+\sin \theta \, \delta S_p,
\]
where
\[
 \delta S_p =
 \frac{1}{16 \pi G }\int_{\mM} \, \mR _{ab}\wedge \theta^{a} \wedge \theta^{b}.
\]
Using the Bianchi identity \eqref{SourcedBianchi} this can be brought to the form
\[
\delta S_p =&
-\frac{1}{16 \pi G }\int_{\mM} \, 
d\left(R_a \wedge \theta^a\right)\\
&+\frac{1}{16 \pi G }\int_{\mM} \, 
R_a \wedge R^a
.
\]
The second term above vanishes because $R_a \wedge R^a=0$. We are left with a total derivative term and therefore the dynamical equations of motion are invariant under the duality operation. Writing the result explicitly in terms of spacetime indices we arrive at
\[
\delta S_p =\frac{1}{16 \pi G } \int_{\mM} d^4x \sqrt{-g} \,\, \nabla_{\mu} D^{\mu},
\]
where
\[
D^{\mu} = \frac{1}{2}\ep^{\mu\nu\rho\sigma}T_{\nu\rho\sigma}
\]
is the gravitational duality current - the analogue of the gauge duality current \eqref{GaugeDualCurrent}.
Here $\ep_{\mu\nu\rho\sigma}$ is the curve-linear Levi-Civita tensor.
To summarize this section, we see that duality invariance can be realized in theories where the matter fields can be accommodated as components of the torsion tensor.

\section{NS-NS Gravity}\label{NSNSgravity}

The field content of the universal massless sector of supergravity, known as NS-NS gravity, contains the metric, dilaton and a Kalb-Ramond two-form.
It is well-known that the torsional degrees of freedom of the Riemann-Cartan manifold can accommodate the NS-NS fields \cite{Scherk:1974mc,Nepomechie:1985us,Monteiro:2021ztt}. The dilaton is assigned to the trace of the contorsion while the B-field is related to its fully antisymmetric component
\[
K_{\nu\rho}{}^{\mu} =
\frac{1}{2\sqrt{3}}H_{\nu\rho}{}^{\mu}
-\frac{1}{2\sqrt{3}} \left(\delta^{\mu}_{\nu}\pa_{\rho}\phi - g_{\nu\rho}g^{\mu\sigma}\pa_{\sigma}\phi\right),
\]
where $H=dB$ is the curvature of the B-field and we follow the notations of \cite{Monteiro:2021ztt}.
The generalized Ricci scalar is then given by
\[
\mR=R -\frac{1}{2}   \nabla_{\mu}\phi\nabla^{\mu}\phi
&-\frac{1}{12}e^{-2\phi}H_{\mu\nu\rho}H^{\mu\nu\rho}
\\&+\sqrt{3}  \, \nabla_{\mu}\nabla^{\mu}\phi.
\]
The NS-NS Lagrangian can therefore be written in terms of the generalized Ricci scalar
\[
S &= \frac{1}{16 \pi G } \int d^4x \, \sqrt{g}\,  \Big(
R
-\frac{1}{2} \, \nabla_{\mu}\phi\nabla^{\mu}\phi
-\frac{1}{12}e^{-2\phi}H_{\mu\nu\rho}H^{\mu\nu\rho}
\Big)\\
&= \frac{1}{16 \pi G } \int d^4x \, \sqrt{g}\, \mR,
\]
namely it is of the Palatini form \eqref{PalatiniAction}.

In four dimensions, the antisymmetric tensor is dual to a pseudo-scalar field $a$, the axion,
\[
H_{\mu\nu\rho} = - e^{ \phi }\ep_{\mu\nu\rho\sigma}\pa^{\sigma}a.
\]
In terms of the axion-dilaton fields, the contorsion is given by
\[\label{NSNScontorsion}
K_{\rho\mu\nu} = -\frac{1}{8\sqrt{3}}
\left(
\delta_{\mu\nu\rho\sigma}\pa^{\sigma}\phi
+\ep_{\mu\nu\rho\sigma}\pa^{\sigma}a
\right),
\]
where
\[
\delta_{\mu\nu\rho\sigma} \equiv \frac{1}{2} \left(
g_{\mu\rho}g_{\nu\sigma}-g_{\mu\sigma}g_{\nu\rho}
\right)
\]
and the NS-NS Lagrangian is
\[\label{NSNSaction}
S  = \frac{1}{16 \pi G } \int d^4x \, \sqrt{g}\,  \left(
R
-\frac{1}{2} \, ( \nabla\phi)^2
-\frac{1}{2} \,  ( \nabla a)^2
\right).
\]
The dynamical equation of motion of the NS-NS action are duality invariant since it is of the Palatini form \eqref{PalatiniAction}, and the duality current is
\[
D^{\mu} = \frac{\sqrt{3}}{4} \pa^{\mu}a.
\]
We see that the gradient of the axion is the generator of duality transformations!
We also observe that axion-dilaton rotation can be implemented by dualizing the contorsion tensor \eqref{NSNScontorsion}.

\section{The Heterotic String}\label{HeteroticString}

The massless bosonic field spectrum of heterotic string theory compactified on a Calabi-Yau manifold \cite{Yau:1977ms,Yau:1978cfy} includes the metric, dilaton, antisymmetric Kalb-Ramond field and one or more gauge potentials.
The low energy action is given by
\[\label{HeteroticAction}
S &= \frac{1}{16 \pi G } \int d^4x \, \sqrt{g}\,  \Big(
R
-\frac{1}{2} \, ( \nabla\phi)^2
-\frac{1}{2} \,  ( \nabla a)^2
\\
&\qquad\qquad\qquad\qquad\quad
-\frac{1}{4}e^{-\phi} F^2
+\frac{1}{4} a\,  F \ast F
\Big).
\]
To leading order in Newton's constant, this matter content can be realized as components of the torsion tensor in the Riemann-Cartan theory. To see this, first we express the gauge fields in terms of the electrostatic $v$ and magnetostatic $u$ potentials
\[
F_{0i} &= \pa_i v ,\\
\Ft_{0i} &=  \pa_i u,
\]
where the index $i=1,2,3$ stands for the spatial coordinates. To leading order in Newton's constant the heterotic action then reduces to
\[
S & \approx \frac{1}{16 \pi G } \int d^4x \, \sqrt{g}\,  \Big(
R
-\frac{1}{2} \, ( \nabla\phi)^2
-\frac{1}{2} \,  ( \nabla a)^2
\\
&\qquad\qquad\qquad\qquad\qquad
-\frac{1}{2} \, ( \nabla v)^2
-\frac{1}{2} \,  ( \nabla u)^2
\Big).
\]
We see that to linear order, the electrostatic and magnetostatic potentials enter into the action in the same way as the dilaton and the axion, respectively.
The linearized generalized Riemann tensor is given by
\[
\mR_{\mu\nu}{}^{\rho\sigma} =&-\frac{1}{2} \pa_{[\mu}\pa^{[\rho}h_{\nu]}{}^{\sigma]}
\\&
+ \frac{1}{4\sqrt{3}} \left(
\delta_{\rho\sigma\lambda[\mu}\pa_{\nu]}\pa^{\lambda} \phi
+\ep_{\rho\sigma\lambda[\mu}\pa_{\nu]}\pa^{\lambda} a
\right)\\
& +\frac{1}{4\sqrt{3}} \left(
\delta_{\rho\sigma\lambda[\mu}\pa_{\nu]}\pa^{\lambda} v
+\ep_{\rho\sigma\lambda[\mu}\pa_{\nu]}\pa^{\lambda} u
\right).
\]
Duality transformation of the generalized Riemann curvature now incorporates three elements at the same time:
\begin{enumerate}
	\item Riemannian duality of the metric $h_{\mu\nu}$,
	\item Axion-dilaton rotation,
	\item Electro-magnetic duality.
\end{enumerate}

\section{Discussion}

In this paper we have explored various aspects of duality in gravitational theories coupled to matter fields. First, we showed that in cases where the matter fields can be understood as components of the torsion tensor, duality can be realized as a symmetry acting on the generalized curvature 2-form. We have computed the corresponding duality current, which is the symmetry generator. As an example, we consider NS-NS gravity, where the duality current was evaluated to be equal to the gradient of the axion field. Finally we considered the low energy limit of the heterotic string compactified on a Calabi-Yau manifold \cite{Yau:1977ms,Yau:1978cfy} and found that in the linearized approximation duality of the generalized Riemann tensor manifest Riemannian, axion-dilaton and electro-magnetic duality all at the same time.

A long-standing problem that we wish to address in the future is how to source the Taub-NUT metric, which is the gravitational analogue of a dyon. Previous works indicate that this can be done using torsion \cite{Kol:2020zth}. We hope that the tools we have developed in this paper will pave the way to solving this problem, for example by dualizing the point-like source of the Schwarzschild metric.

The double copy structure of scattering amplitudes in the linearized NS-NS theory, as well as in the heterotic string, was explored in \cite{Monteiro:2021ztt}.
Beyond the linear approximation, it would be interesting to study exact solutions of the heterotic string such as the solutions of the Strominger system \cite{Strominger:1986uh,Strominger:1990et,Li:2004hx}.
In particular, we would like to understand if duality and the double copy structure are manifested in the Strominger system.
Various related aspects of the double copy, S-duality and connections to Ehlers/Geroch transformations were studied in \cite{Luna:2015paa,Huang:2019cja,Emond:2020lwi,Alawadhi:2019urr,Banerjee:2019saj}.

Finally, the authors of \cite{Seraj:2021rxd,Seraj:2022qyt} have studied the gyroscopic gravitational memory effect and showed that it receives contributions from the generator of electric-magnetic duality on the asymptotic phase space. It would be interesting to understand the relation between these results and our non-linear realization of duality.

\subsection{Acknowledgments}

UK is supported by the Center for Mathematical Sciences and Applications at Harvard University.



\bibliography{bibliography}
\bibliographystyle{apsrev4-2.bst}

\end{document}